\documentclass[aps,prl,twocolumn,groupedaddress,showpacs]{revtex4}

\usepackage[dvips]{graphicx,color}
\begin{document}


\title{Observation of Heteronuclear Feshbach Resonances in a Bose-Fermi Mixture}


\author{S. Inouye}
\author{J. Goldwin}
\author{M. L. Olsen}
\author{C. Ticknor}
\author{J. L. Bohn}
\author{D. S. Jin}
\altaffiliation{Quantum Physics Division, National Institute of
Standards and Technology}

\affiliation{JILA, National Institute of Standards and Technology
and University of Colorado, Boulder, Colorado 80309}


\date{\today}

\begin{abstract}
Three magnetic-field induced heteronuclear Feshbach resonances
were identified in collisions between bosonic $^{87}$Rb and
fermionic $^{40}$K atoms in their absolute ground states.  Strong
inelastic loss from an optically trapped mixture was observed at
the resonance positions of 492, 512, and 543 $\pm$ 2 G. The
magnetic-field locations of these resonances place a tight
constraint on the triplet and singlet cross-species scattering
lengths, yielding $-281\pm 15\, a_0$ and $-54 \pm 12\, a_0$,
respectively. The width of the loss feature at 543 G is $3.7 \pm
1.5$ G wide; this broad Feshbach resonance should enable
experimental control of the interspecies interactions.
\end{abstract}

\pacs{34.50.-s}

\maketitle

Since their first observations in 1998~\cite{ShinFR},
magnetic-field Feshbach resonances in ultracold
collisions~\cite{FRtheory} have been used as a versatile tool for
manipulating quantum degenerate atomic gases. Simply by varying
the strength of an applied magnetic field, experimenters can
control the collisional interactions between ultracold atoms. The
scattering length, which characterizes these interactions, can be
tuned from positive infinity through zero to negative infinity.
The unique tunability provided by Feshbach resonances has enabled
the controlled collapse of a Bose-Einstein condensate
(BEC)~\cite{stable85Rb}, the creation of bright matter wave
solitons~\cite{solitons}, the formation of ultracold diatomic
molecules~\cite{molecules}, and the realization of the BCS--BEC
crossover in dilute gases~\cite{BCS-BEC}. Although Feshbach
resonances have now been observed for many of the alkali atoms,
including both bosonic and fermionic species, interspecies
resonances involving heteronuclear collisions have not yet been
observed~\cite{note}.  As pointed out in Ref. \cite{LENSFR}, an
interspecies Feshbach resonance would open up new possibilities
such as boson-mediated Cooper pairing~\cite{BFBCS} and the
creation of ultracold polar molecules~\cite{polarmolecules}.

In this Letter we report the first observation of Feshbach
resonances in the scattering between two distinct species,
$^{87}$Rb (boson) and $^{40}$K (fermion). We have located three
interspecies resonances between $^{87}$Rb atoms in the
$|F,m_F\rangle = |1,1\rangle$ state and $^{40}$K atoms in $|9/2,
-9/2\rangle$ states.  Here $F$ is the total spin, and $m_F$ is the
spin projection. Resonances at 492, 512, and 543 $\pm$ 2 G were
located from measurements of inelastic loss in a search that
initially covered the range from 18 G to 635 G. The widest of the
features we observe has a full-width at half-maximum (FWHM) of
$3.7\pm 1.5$ G; this resonance is sufficiently broad in magnetic field to
permit ready experimental control over the interspecies
interactions in this Bose-Fermi quantum gas mixture.

The observation of magnetic-field Feshbach resonances can also
play an important role in improving knowledge of the interatomic
molecular potentials, often to an unprecedented level of
accuracy~\cite{FRspectroscopy,RbFR}. Such accuracy is possible
because the location of the Feshbach resonance is very sensitive
to the energies of the most weakly bound molecular states. For the
case of heteronuclear collisions, in particular, there is often a
lack of other spectroscopic data, such as two-color
photoassociation spectra, that probe the long-range part of the
interatomic potential. In the case of $^{87}$Rb and $^{40}$K,
nonresonant collision measurements have provided information about
the scattering parameters
\cite{LENScollsOlder,LENScollsNewer,announcement} and enabled a
prediction for Feshbach resonances in this system \cite{LENSFR}.
We compare the observed magnetic-field locations of the
$^{87}$Rb-$^{40}$K resonances reported here with this prediction,
and from our data provide an improved determination of both the
triplet and singlet scattering lengths.

The details of our apparatus and cooling scheme are presented in
Ref. \cite{announcement}, and will only be briefly recounted here.
We simultaneously laser cool $^{87}$Rb and $^{40}$K atoms in a
two-species magneto-optical trap \cite{MOT}.  The $^{87}$Rb
($^{40}$K) atoms are optically pumped to the $|2, 2\rangle$
($|9/2,9/2\rangle$) state, and captured in a quadrupole magnetic
trap mounted on a motorized translation stage.  The trap is then
physically transported to an ultra-high vacuum cell, where the
atoms are transferred into an Ioffe-Pritchard type magnetic trap
for forced radio-frequency (rf) evaporative cooling.  The
$^{87}$Rb gas is selectively evaporated, with the $^{40}$K gas
cooled sympathetically through thermal contact with the Bose gas.
At the end of the cooling cycle, a $^{87}$Rb BEC coexists with a
degenerate Fermi gas of $^{40}$K atoms at temperatures down to
$0.2\, T_F$, where $T_F$ is the Fermi temperature. For the
measurements reported here, however, the evaporation is stopped
prior to achieving quantum degeneracy to minimize inelastic losses
due to the high density of the $^{87}$Rb BEC and prevent the
possibility of mechanical collapse of the mixture
\cite{LENScollapse}.

\begin{figure}
\includegraphics[scale=0.7]{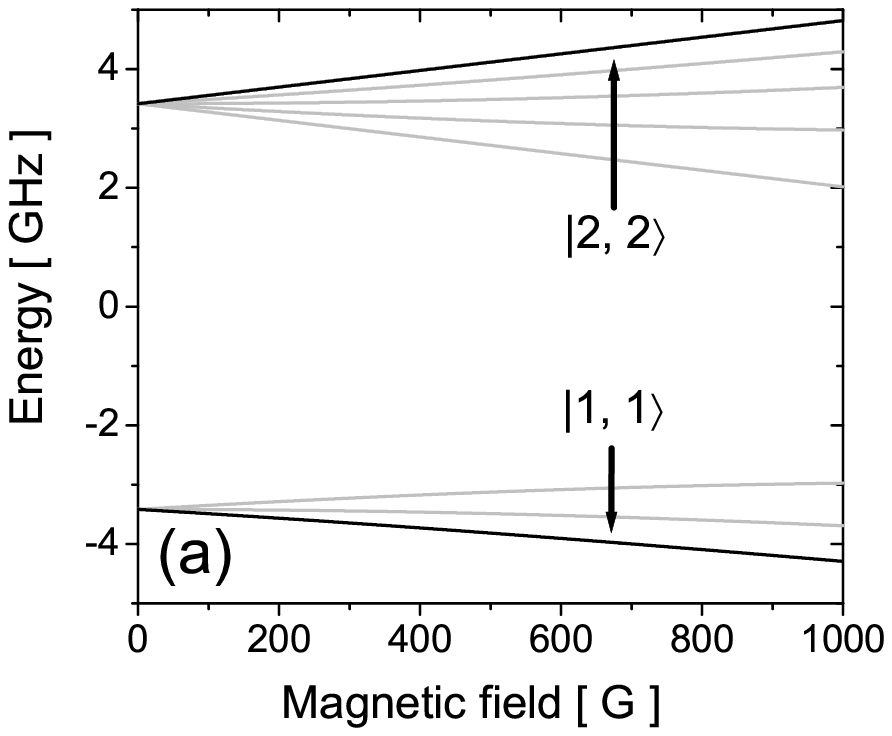}\hfill\includegraphics[scale=0.7]{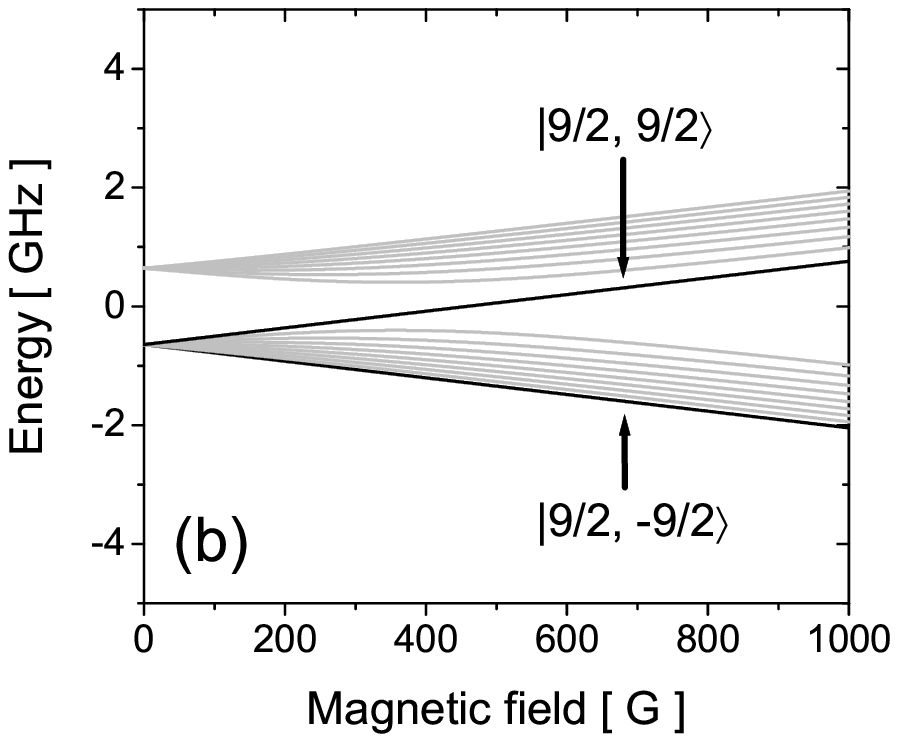}
\caption{(a) Hyperfine states of $^{87}$Rb as a function of
magnetic field $B$. The solid black line in the upper ground state
manifold is the $|2,2\rangle$ state where we perform the
evaporative cooling in the magnetic trap. Atoms are transferred to
the absolute ground state ($|1,1\rangle$ state; the solid black
line in the lower manifold) by applying a frequency-swept
microwave field. (b) Hyperfine states of $^{40}$K as a function of
magnetic field. The stretched state ($|9/2,9/2\rangle$; the upper
solid black line) used for sympathetic cooling resides in the
lower manifold. Atoms are transferred to the absolute ground state
($|9/2, -9/2\rangle$; the lower solid black line) by driving a
multi-level adiabatic rapid passage induced by a frequency-swept
rf field. \label{fig:Breit-Rabi}}
\end{figure}

To avoid any complications from dipolar collisional loss, which is
typically magnetic-field dependent, we put the atoms in the lowest
energy magnetic sublevels for searching for Feshbach resonances
(Fig.\ref{fig:Breit-Rabi}). Since these states are high-field
seeking, and therefore not confined in our magnetic trap, we use a
far off-resonance optical dipole trap (FORT) to confine the gas
mixture.  We load $3\times 10^5$ Rb atoms and $3\times 10^4$ K
atoms into the FORT formed at the focus of a Yb:YAG laser beam.
The laser operates at a wavelength $\lambda= 1030$ nm and at the
focus the intensity profile has a $1/e^2$ radius of $20.5$ $\mu$m.
For loading, the laser power is linearly increased to 1 W over 500
ms, and then the magnetic trap is shut off. A 3 G bias field
remains in order to maintain the spin polarization of the atoms.
At the end of the FORT loading sequence, the temperature of the
mixture is $14\,\mu$K. For comparison, at 1 W the calculated
optical trap depth is $\sim 200\,\mu$K$\times\, k_B$, where $k_B$
is Boltzmann's constant, and the trap has a radial (axial)
trapping frequency for $^{87}$Rb atoms of 2.3 kHz (24 Hz).

After loading the optical trap, $^{87}$Rb atoms are transferred from the
$|2,2\rangle$ state to the $|1,1\rangle$ state via adiabatic rapid
passage with a 20 ms frequency sweep of an applied microwave
field. The efficiency of the transfer is better than 90\%, and the
remaining $|2,2\rangle$ atoms are immediately removed from the
trap by a 5 ms pulse of light resonant with the $F=2\rightarrow
F^\prime=3$ transition. The magnetic field is then increased to 18
G in 100 ms, and the K atoms are transferred from the
$|9/2,9/2\rangle$ state to the $|9/2,-9/2\rangle$ state via
adiabatic rapid passage induced by an rf field that is frequency
swept across the ten magnetic sublevels \cite{CindyFR}. The
magnetic field could then be increased to as high as 635 G in
order to search for magnetic-field Feshbach resonances. At our
highest fields we estimate the magnetic-field difference across
the long dimension of the cloud to be less than 10 mG.

To search for Feshbach resonances, we have looked for the enhanced
inelastic loss that is common in their vicinities\cite{ShinFR}.
Data were taken by quickly (in 10 ms) increasing the field to some
value and then applying a slow (1.08 s duration) magnetic-field
sweep over a limited range to look for loss features. After
returning to the low magnetic field (in 10 ms), the number of
remaining $^{40}$K atoms was determined from resonant absorption
images taken in the optical trap. Initial data covered magnetic
fields from 18 G to 635 G using 30 G sweeps. Subsequent data
zoomed in on the only region where loss was observed.  Finally
having distinguished three loss features, data was taken holding
the magnetic field at a constant value $B$ for the entire 1.08 s
to more precisely determine their magnetic-field locations and
widths.

\begin{figure}
\includegraphics[scale=0.45]{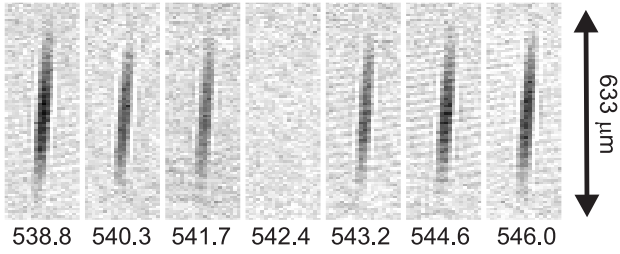}
\caption{In-trap absorption images of $^{40}$K after holding 1.08
s at various magnetic fields in the vicinity of an interspecies
Feshbach resonance with $^{87}$Rb atoms.  The label on each figure
gives the magnetic field in Gauss.  There is a systematic $\pm 2$
G uncertainty on the magnetic-field value from our calibration. No
$^{40}$K atoms could be seen after holding at 542.4 G.
\label{fig:images}}
\end{figure}

\begin{figure}
\includegraphics[scale=0.7]{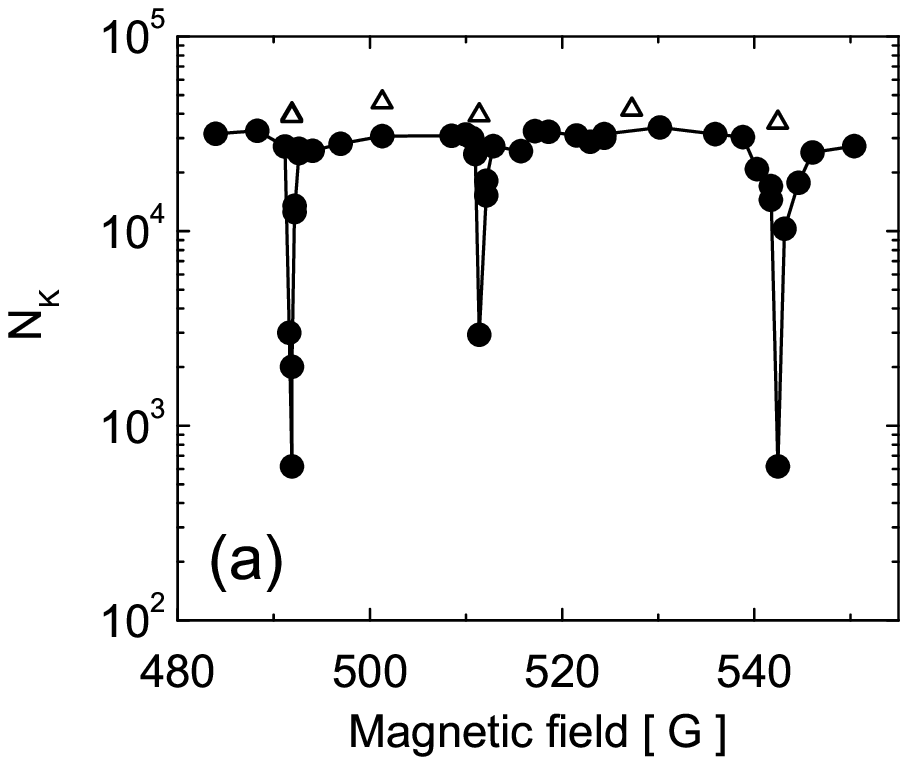}\hfill\includegraphics[scale=0.7]{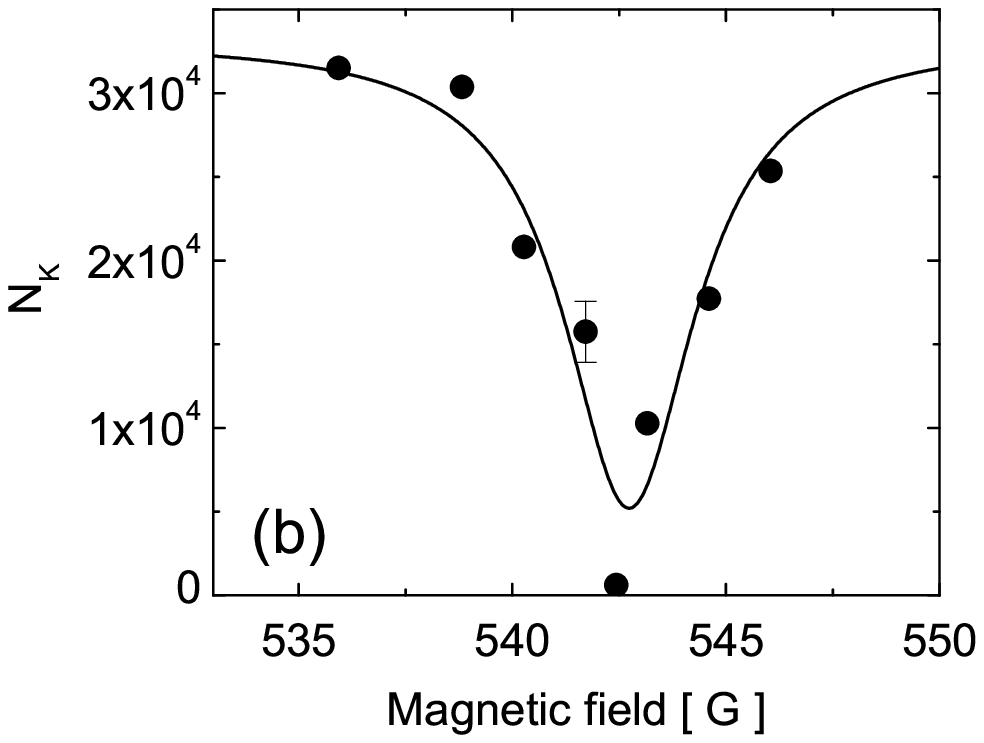}
\caption{Observation of the three inter-species Feshbach
resonances. (a) Inelastic loss of $^{40}$K atoms in the
$|9/2,-9/2\rangle$ state was measured by holding the
$^{87}$Rb-$^{40}$K mixture in the optical dipole trap for 1.08 s
in a fixed magnetic field $B$. The number of remaining $^{40}$K
atoms (shown in filled circles) shows three narrow features as a
function of $B$. The loss features were not observed when
$^{87}$Rb atoms were removed from the mixture (empty triangles).
(b) Close-up of the highest field resonance. The solid line is a
Lorentzian fit to the loss with FWHM of $3.7 \pm 1.5$ G. The error
bar reflects only the statistical uncertainty.
\label{fig:resonances}}
\end{figure}

Figure \ref{fig:images} shows in-trap absorption images of
$^{40}$K after holding the $^{87}$Rb-$^{40}$K gas mixture at
constant $B$ in the vicinity of the broadest observed resonance.
The reduced number of $^{40}$K atoms near $B=542.4$ G can clearly
be seen as a decreased optical depth of the trapped gas. Figure
\ref{fig:resonances} shows the measured number of $^{40}$K atoms
remaining as a function of $B$. We observe three distinct loss
features which we interpret as resulting from enhanced rates for
three-body inelastic collisions near inter-species Feshbach
resonances. One would also expect a corresponding loss of Rb atoms
from the mixture.  However, this was not discernable with our 10:1
ratio of Rb to K atoms.  To verify that the K loss features depend
on interspecies collisions, we repeated the measurements adding a
complete removal of the $^{87}$Rb atoms by forced evaporation
prior to loading the FORT. For this case, we do not observe any
loss in the single-species $^{40}$K gas at any $B$.

The loss features occur at 492, 512, and 543 $\pm$ 2 G, where the
uncertainty comes from a systematic uncertainty in our calibration
of $B$. The widths are measured to be $0.7\pm 0.2$, $0.4\pm 0.2$,
and $3.7\pm 1.5$ G, respectively. However, for the two lowest
field resonances the measured width may be limited by imperfect
magnetic field stability over the 1.08 sec measurement. The
observed 3.7 G width of the high-field feature, on the other hand,
suggests a Feshbach resonance that is sufficiently broad to enable
a wide variety of experiments. Furthermore, the fact that we could
observe such narrow features with a one-second hold, which is long
compared to any dynamics of the trapped gas, suggests that
inelastic losses near the resonances should be easily managed
during future experiments. Finally we note that although we did
not observe any other loss features between 18 and 635 G, we
cannot rule out the presence of other more narrow resonances, or
resonances with highly suppressed inelastic losses.

\begin{figure}
\includegraphics[scale=0.7]{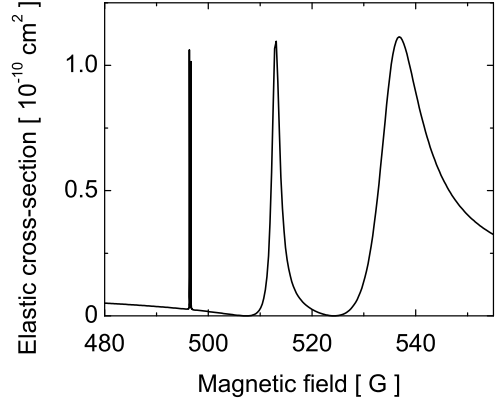}
\caption{Calculated elastic cross section for a fixed collision
energy of 10 $\mu$K between $^{87}$Rb atoms in the $|1,1\rangle$
state and $^{40}$K atoms in the $|9/2,-9/2\rangle$ state as a
function of magnetic field. The triplet (singlet) scattering
length used for producing this result was $-281\, a_0$ ($-54\, a_0
$). \label{fig:sigma}}
\end{figure}

Figure \ref{fig:sigma} shows the elastic cross section for
collisions between $|9/2,-9/2\rangle$ $^{40}$K atoms and
$|1,1\rangle$ $^{87}$Rb atoms, calculated using the parameters
obtained from the position of the three observed Feshbach
resonances. We expect the inelastic loss peaks to coincide with
the elastic peaks   to within the experimental uncertainty
\cite{pwave}. The scattering Hamiltonian was constructed in the
field-dressed hyperfine basis, by standard methods \cite{Burke}.
This Hamiltonian uses the {\it ab initio} singlet and triplet
potentials from Ref. \cite{Ross}, smoothly matched to the
long-range dispersion potentials of the form
$V_{long}=-C_6/R^6-C_8/R^8-C_{10}/R^{10}\pm V_{ex}$. The $C_6$
coefficient is given by Ref. \cite{Der} and the $C_8$ and $C_{10}$
coefficients are found in Ref. \cite{Mar}.  The exchange potential
$V_{ex}$ is estimated using the analytic form of Ref. \cite{Ex}.
These potentials are consistent with the ones used in the
calculations of Ref. \cite{LENScollsOlder,LENSFR}. In addition,
the singlet and triplet potentials can be varied at short
internuclear separation (shorter than their equilibrium
separation), to allow fine tuning of their scattering lengths.

To fit the data, the singlet and triplet scattering lengths ($a_s$
and $a_t$, respectively) were varied until all three resonances
corresponded with the experimentally measured positions.  This
analysis demonstrates that the two higher-field resonances are
s-wave in character, while the narrow feature at 492 Gauss
represents a $p$-wave resonance.  This resonance was predicted in
Ref. \cite{Ticknor}, based on earlier estimates of scattering
lengths. The position of a narrow nearby $s$-wave resonance,
predicted in Ref. \cite{LENSFR}, is refined by this analysis; we
now expect this resonance to lie at 444 G. The result of this
assignment of the features establishes the interspecies scattering
lengths to be $a_t=-281 \pm 15\, a_0$ and $a_s=-54 \pm 12\, a_0$.

The resulting values for the triplet and singlet scattering
lengths are in good agreement with collisional measurements of the
elastic collision cross section
\cite{LENSFR,LENScollsOlder,announcement}. Note that there is a
serious disagreement when comparing our result with the value of
$a_t = -395\pm 15\, a_0$ determined from comparing theory
predictions and experimental observation of collapse phenomena
\cite{modu03}. These results suggest a need for further
investigation of the behavior of the mixture close to the
collapse.  The control over interactions afforded by an
interspecies Feshbach resonance should greatly facilitate studies
of such interaction-driven phenomena.

In conclusion, we have observed three inter-species Feshbach
resonances in a Bose-Fermi mixture.  The resonances were located
by observing inelastic loss of $^{40}$K atoms from a
$^{87}$Rb-$^{40}$K mixture in the lowest energy spin states. The
location and widths of the loss features are in reasonable
agreement with a recent prediction \cite{LENSFR} and enable us to
make a more precise determination of the singlet and triplet
scattering lengths.  Future work exploiting these Feshbach
resonances to control the interactions in a Bose-Fermi mixture
opens up a number of exciting possibilities such as controlled
collapse due to mechanical instability, creation of heteronuclear,
fermionic molecules with anisotropic dipole-dipole interactions,
fermion-mediated bright solitons in a BEC \cite{BFsolitons},
complex phase diagrams in optical lattices \cite{Zollerpaper}, and
a proposed $p$-wave Cooper pairing mechanism \cite{BFBCS}, where
the effective attraction between fermions is generated by a mutual
interaction with phonons in the condensate.

\begin{acknowledgments}
This work was funded by a grant from the U. S. Department of
Energy, Office of Basic Energy Sciences, and the National Science
Foundation. We also acknowledge useful discussions with the
members of the JILA quantum gas collaboration.  The computing
cluster used for these calculations was provided by a grant from
the W. M. Keck Foundation.
\end{acknowledgments}

\bibliography{basename of .bib file}

\newpage

\newpage

\newpage

\newpage

\end{document}